# A meta-generalized gradient approximation-based time-dependent and dielectric function dependent method for optical properties of solid materials

Hong Tang[1], Niraj Pangeni, and Adrienn Ruzsinszky

Department of Physics and Engineering Physics, Tulane University, New Orleans, LA 70118

**ABSTRACT**

Accurate and efficient calculation of optical response properties of solid materials is still challenging. We present a meta-generalized gradient approximation (metaGGA) density functional based time-dependent and dielectric function dependent method for calculating optical absorption, exciton binding energy and intrinsic exciton lifetime for bulk solids and two-dimensional (2D) monolayer materials. This method uses advanced metaGGA functionals to describe the band structures, and a dielectric function mBSE (model Bethe-Salpeter equation) to capture the screening effect accurately and efficiently and the interaction between electrons and holes. The calculated optical absorption spectra of bulk Si, diamond, SiC, MgO, and monolayer $MoS_2$ qualitatively agree with experimental results. The exciton binding energies of the first prominent peak in the optical absorption spectra of the direct band gap solids Ar, NaCl and MgO from mBSE qualitatively agree with those from standard GW-BSE. For monolayer $MoS_2$, mBSE predicts quantitatively accurate binding energy for the first prominent peak, better than GW-BSE does. The calculated intrinsic exciton lifetimes for materials considered here show magnitudes of several nanoseconds for most bright excitons. The presented mtaGGA-mBSE method is established as a computationally efficient alternative for optical properties of materials with an overall qualitative accuracy.

[1] Contact email: htang5@tulane.edu

## I. INTRODUCTION

Calculations of optical response properties, such as optical absorption, exciton binding energy, etc., are important fingerprints of the nature of materials, and guide us for designing and engineering novel materials and devices [1-3]. The standard and generally accurate approach to calculate optical properties of semiconducting materials is the many-body perturbation method GW and BSE (Bethe-Salpeter equation) [4, 5]. The bottleneck of this method is its high computational cost. The conventionally used one-shot $G_0W_0$ method for quasiparticle energy calculation shows starting point dependence on density functional approximations (DFA), and the self-consistent GW, comes with even higher computational cost, and also suffers from the lack of correct vertex corrections [6].

An alternative more feasible approach for optical response calculation, is linear-response time dependent density functional theory (TDDFT) [7-9], which is based on the time-dependent Kohn-Sham equation with a time-dependent exchange-correlation potential Vxc, and an exchange-correlation kernel fxc. The kernel is the second functional derivative of the exchange-correlation energy with respect to the time-dependent density within the adiabatic approach. fxc is essential since it represents the effective interaction between electrons in the system. In practice, analogous to ground state DFT, both Vxc and fxc must be approximated. Several recent approximations for fxc have demonstrated already improved accuracy for optical absorption and excitonic effects [10-13].

In TDDFT, the electron energies for the band structure and the wavefunctions can be obtained from semilocal density functional approximations (DFAs) or hybrid functionals. While conventional DFAs underestimate the band gaps for semiconductors, hybrid functionals with proper inclusion of the nonlocal exact exchange and screening can improve the description for the band gap. Recently, a nonempirical hybrid function scheme, called the dielectric function dependent hybrid (DDH) [14, 15], has been proposed. This approach can compete with GW for fundamental band gaps, while DDH within TDDFT (TD-DDH) can reach the accuracy of GW-BSE for optical absorption, as confirmed for bulk solids [15]. The accuracy of TD-DDH for 2D materials, however, remains to be tested. In the TD-DDH method, a model of the static dielectric function as a function of wave vector q is approximated and parameter-fitted to GW results. This model dielectric function can properly account for the wave-vector-dependent screening, with a weakly screened exchange interaction at the short range and a dielectric function reduced exchange interaction in the long range.

MetaGGA (meta-generalized gradient approximation) density functionals [16, 17] represent a higher level than conventional semilocal density functionals, such as LDA (local density approximation) [18] and GGAs (generalized gradient approximations) [19]. Kinetic energy density ($\tau$) dependent metaGGA functionals, such as the strongly constrained and appropriately normed (SCAN) [16] and (modified TASK) mTASK [17], are different great choices for the ground state or for the fundamental band gap. SCAN and mTASK are both different from LDA and GGAs with greater potential. With the occupation number $f_i$, the kinetic energy density $\tau(\mathbf{r})$ ( $\tau(\mathbf{r}) = \frac{1}{2}\sum_i f_i |\nabla \psi_i(\mathbf{r})|^2$ ) involves all the orbitals. Further ingredients derived from $\tau(\mathbf{r})$, such as the dimensionless deviation from the single orbitals shape $\alpha(\mathbf{r})$ [16], leave a vast platform of physical scenarios to be properly described by these metaGGAs. The $\tau(\mathbf{r})$ dependence makes metaGGAs mimic nonlocal exact exchange effects to some extent. The same spatial nonlocality can also be harnessed within TDDFT [20]. MetaGGAs' computational cost is about the same level as that of conventional GGA functionals, but much less than that of hybrid functionals. The mTASK functional, with an enhanced derivative of exchange energy density with respect to $\tau$, can predict the correct band gaps for many semiconducting and insulating solids including low dimensional layered materials [17].

In this work, we combine the metaGGA and DDH scheme and establish this combination as an emerging alternative to GW-BSE for the optical response calculations. The metaGGA-DDH framework is applied in the spirit of TD-DDH [15], for a range of materials, including the bulk solids Si, diamond, SiC, NaCl, Ar and MgO, and low-dimensional prototype materials hBN and $MoS_2$ monolayers. We found that the method can achieve an accuracy close to that of the hybrid-based DDH, however, with much less computational cost, providing an efficient computational method for optical response properties of solid materials.

## II. COMPUTATIONAL METHODS

The calculations were conducted in the Vienna Ab initio Software Package (VASP) [21] with projector augmented-wave (PAW) pseudopotentials [22]. For the diamond cubic structure solids Si, diamond and SiC, the lattice constants are set to 5.422 [23], 3.555 [23] and 4.348 Å [23], respectively. For the rock salt structure solids Ar, NaCl and MgO, the lattice constants are set to 5.311 [24], 5.547 [25] and 4.212 Å [26], respectively. For the two-dimensional (2D) hexagonal monolayers hBN and $MoS_2$, the in-plane lattice constants are set to 2.511 [27] and 3.174 Å [28], respectively. The lattice constants are set to either experimental values or the SCAN relaxed values, which are very close to the experimental values. For all materials, the atomic coordinates are fully relaxed with SCAN with the force on each atom less than 0.01 eV/Å. The k-grid mesh is $12 \times 12 \times 12$ for Si, Ar and NaCl, $15 \times 15 \times 15$ for diamond, SiC and MgO, $20 \times 20 \times 1$ for hBN and is $18 \times 18 \times 1$ for $MoS_2$. The energy cutoff for plane waves is 480 eV. Four valence bands and four conduction bands are set to optical calculations.

For optical calculations, our metaGGA based TDDFT approximation resembles the model BSE (mBSE) method [14, 15] in the construction of the screening. mBSE and TDDFT are both based on the Casida equation [29] formalism. We build the mBSE screening based on the dielectric function from $G_0W_0$@SCAN, where the static inverse dielectric function $1/\varepsilon_\infty(q)$ as a function of wave vector $q$ can be obtained. We believe that the computational expense of the fitting to $G_0W_0$@SCAN pays off for low dimensional materials where conventional computational techniques can potentially break down. Then, two screening parameters $\alpha$ and $\mu$ are obtained from a fitting of the static dielectric function with the following model [15],

$$\varepsilon^{-1}(q) = 1 - (1-\alpha)\exp\left(-\frac{q^2}{4\mu^2}\right). \qquad (1)$$

Figure 1 shows the calculated inverse static dielectric function as a function of wavevector and the fitting curves to Eq. (1) for the six bulk solids Si, diamond, SiC, Ar, NaCl and MgO, and two 2D monolayers hBN and MoS$_2$ considered in the work. The obtained two screening parameters $\alpha$ and $\mu$ are listed in Table I. In our mBSE approach, for computational saving we keep only the diagonal elements of the corresponding dielectric matrix, ignore the frequency dependence, and model the effect of the off-diagonal matrix elements or local-field effects by a short-ranged static exchange-correlation kernel $f_{xc}^{loc}$. We denote the methods as SCAN-mBSE-$f_{xc}^{loc}$ and mTASK-mBSE-$f_{xc}^{loc}$. The semilocal exchange and correlation potentials are built from SCAN and mTASK, respectively.

In TDDFT, the coupling matrix F$_{\text{Hxc}}$ in the Casida equation can be written as the sum of three contributions, namely, the Hartree term, the screened exchange term $W^m$ and the local exchange-correlation term $f_{xc}^{loc}$ [15]. In mBSE or DDH [15], the screened exchange term $W^m$ is included and $W^m$ is calculated as the product of the dielectric function model (Eq. (1)) and the bare Coulomb interaction. Within this framework, we replace the PBE GGA by a meta-GGA. The standard BSE employs a static dielectric *matrix* that is computed and not modelled.

In analogy to DDH [15], we replace the GW exchange-correlation self-energy (used in standard GW+BSE) by the meta-GGA hybrid exchange-correlation potential $V_{xc}$ in our SCAN-mBSE-$f_{xc}^{loc}$ and mTASK-mBSE-$f_{xc}^{loc}$ methods as

$$V_{xc}(\mathbf{r},\mathbf{r}') = [1-(1-\varepsilon_\infty^{-1})\operatorname{erf}(\mu|\mathbf{r}-\mathbf{r}'|)]V_x^{Fock}(\mathbf{r},\mathbf{r}') + (1-\varepsilon_\infty^{-1})V_x^{local}(\mathbf{r};\mu)\delta(\mathbf{r}-\mathbf{r}') + V_c^{local}(\mathbf{r})\delta(\mathbf{r}-\mathbf{r}'), \qquad (2)$$

where $V_x^{Fock}$ is the Fock exchange. $V_x^{local}$ and $V_c^{local}$ are the local parts of exchange and correlation potentials and calculated from metaGGA-based electron density and wavefunctions. The local exchange-correlation interaction $f_{xc}^{loc}$ is expressed as

$$f_{xc}^{loc}(\mathbf{r},\mathbf{r}') = \frac{\delta\{V_c^{local}+(1-\varepsilon_\infty^{-1})V_x^{local}\}}{\delta\rho(\mathbf{r})}\delta(\mathbf{r}-\mathbf{r}'), \qquad (3)$$

where $\rho(\mathbf{r})$ is the electron density. $f_{xc}^{loc}$ is built up from SCAN or mTASK as well. Note that in equation (3) only functional derivatives of $\rho(\mathbf{r})$ and $\nabla\rho(\mathbf{r})$ are evaluated. MetaGGAs are implicit functions of density. MetaGGA-based kernels rigorously derived with the kinetic energy density dependence are a promising direction for excitonic effects [20, 30], but the derivation of a full metaGGA-based kernel is currently beyond the scope of this work.

### III. RESULTS AND DISCUSSION

### A. Optical absorption spectrum

Table II shows the fundamental band gaps of bulk Si, diamond, SiC, Ar, NaCl and MgO, and 2D monolayers hBN and MoS$_2$ calculated with SCAN and mTASK, in comparison with the GW, TD-DDH [15] and experimental values., SCAN generally underestimates the band gap. mTASK, with an improved slope $\frac{\partial \varepsilon_x}{\partial \alpha}$ and nonlocality [17], where $\varepsilon_x$ is the electron exchange energy density that can distinguish covalent, metallic and vdW bonds, improves band gap predictions. Band gaps from mTASK are approximately as good as those of the computationally costly GW and TD-DDH.

Figure 2 shows the optical absorption spectra calculated with SCAN-mBSE-$f_{xc}^{loc}$ and mTASK-mBSE-$f_{xc}^{loc}$ in comparison with the experimental or GW-BSE results for bulk Si, diamond, SiC, Ar, NaCl and MgO, and for the 2D monolayers hBN and MoS$_2$. The vertical axis represents the imaginary part of the mBSE calculated macroscopic dielectric function. Since SCAN underestimates the fundamental band gaps for these materials, a scissor correction for band gaps is used in the SCAN-mBSE-$f_{xc}^{loc}$ calculation. For the more spatially nonlocal mTASK-mBSE-$f_{xc}^{loc}$, no scissor correction is used for the band gap.

For bulk Si, as seen in Figure 2a, SCAN-mBSE-$f_{xc}^{loc}$ gives the first peak position at 3.37 eV, very close to the experimental value of 3.44 eV. The height of the first peak calculated by SCAN-mBSE-$f_{xc}^{loc}$ is moderately underestimated. SCAN-mBSE-$f_{xc}^{loc}$ gives the second peak position at 4.16 eV, very close to the experimental value at 4.24 eV. The height of the second peak calculated by SCAN-mBSE-$f_{xc}^{loc}$ agrees reasonably with the experimental one. Approximately speaking, the SCAN-mBSE-$f_{xc}^{loc}$ calculated optical absorption spectrum of bulk Si matches the experiment very well, close to or slightly better than the quality of the TD-DDH and GW-BSE results [15]. mTASK predicts 1.33 eV for the band gap of bulk Si, very close to the experimental gap of 1.23 eV. From the energy range of 3.0-4.5 eV, the optical absorption spectrum calculated from mTASK-mBSE-$f_{xc}^{loc}$ shows an appearance of small peaks and shoulders, while at 4.8 eV and 5.67 eV, showing two prominent peaks. Overall, the optical absorption spectrum from mTASK-mBSE-$f_{xc}^{loc}$ doesn't align very well with the experimental one.

For bulk diamond and SiC as shown in Figures 2b and c, the experimental optical absorption shows a smooth hump or peak at about 12 eV and 7 eV, respectively. For diamond, the calculated absorption curves from SCAN-mBSE-$f_{xc}^{loc}$ and mTASK-mBSE-$f_{xc}^{loc}$ are similar. Both methods show a wiggly multi-peak feature from 6-11 eV. The position of the prominent peak calculated from SCAN-mBSE-$f_{xc}^{loc}$ is at about 13 eV, higher than the prediction from mTASK-mBSE-$f_{xc}^{loc}$ at about 12 eV. The positions of the prominent peak calculated from TD-DDH and GW-BSE are also overestimated by about 1 eV [15]. Approximately speaking, for diamond, both SCAN-mBSE-$f_{xc}^{loc}$ and mTASK-mBSE-$f_{xc}^{loc}$ give a qualitatively similar optical absorption curve, more similar to each? other than to the experimental result. For bulk SiC, the shape of the absorption curves calculated from SCAN-mBSE-$f_{xc}^{loc}$ and mTASK-mBSE-$f_{xc}^{loc}$ are like each other, and both resemble the experimental one. The calculated curves show a slightly lower peak height and an overestimated (by about 1.5 eV) peak position.

For the Ar solid shown in Figure 2d, the experimental optical absorption spectrum exhibits two sharp peaks close to each other at about 12 eV. The GW-BSE and TD-DDH methods [15] produce a peak with a much lower height at 11.73 eV and 12.79 eV, respectively. SCAN-mBSE-$f_{xc}^{loc}$ produces a peak at 12.84 eV with a much-reduced height, in good agreement with TD-DDH [15]. The curve calculated by mTASK-mBSE-$f_{xc}^{loc}$ has a peak at 13.3 eV with a very low height, which is hardly discernible in Figure 2d. Some humps also appear within 16-18 eV.

For NaCl shown in Figure 2e, again, two sharp very-close-to-each-other peaks are at about 8.0 eV in the experimental optical absorption spectrum. This feature is not reproduced by SCAN-mBSE-$f_{xc}^{loc}$ and mTASK-mBSE-$f_{xc}^{loc}$, both showing very low peak heights in this energy range. For MgO shown in Figure 2f, the experimental spectrum shows a sharp peak at 7.51 eV. SCAN-mBSE-$f_{xc}^{loc}$ predicts a sharp peak near this energy with a slightly overestimated energy at 7.91 eV. However, mTASK-mBSE-$f_{xc}^{loc}$ gives a peak at 7.6 eV with a very low height. Both GW-BSE and TD-DDH [15] predict a peak with a much-reduced height around 8.0 eV. The experimental spectrum feature within 8-14 eV is approximately reproduced by GW-BSE and TD-DDH [15], while SCAN-mBSE-$f_{xc}^{loc}$ and mTASK-mBSE-$f_{xc}^{loc}$ produce the curves much lower than the experimental one within this energy range.

For monolayer hBN as shown in Figure 2g, the experimental optical absorption spectrum is lacking. However, the GW-BSE spectrum [38] is expected be close to the experimental one, since the experimentally determined optical gap of monolayer hBN is about 6.1 eV [39], close to the first peak position calculated from GW-BSE. SCAN-mBSE-$f_{xc}^{loc}$ produces a sharp peak shifted to a much lower energy 3.41 eV. The mTASK-mBSE-$f_{xc}^{loc}$ produced optical absorption spectrum has a very low absorption below 6.5 eV. For monolayer MoS$_2$, the experimental curve shows two absorption peaks around 2 eV. This feature is reproduced by SCAN-mBSE-$f_{xc}^{loc}$, while mTASK-mBSE-$f_{xc}^{loc}$ produces two much-low-height peaks at slightly lower energies around the energy range. Above 2.5 eV, the experimental spectrum displays much higher absorptions, related to electron-phonon coupling effects [40], which are absent in our current calculation. The spectra from SCAN-mBSE-$f_{xc}^{loc}$ and mTASK-mBSE-$f_{xc}^{loc}$ in this energy range are much lower than the experimental one.

## B. Exciton binding energies

The bulk solids Ar, NaCl, MgO and monolayers hBN and MoS$_2$ are insulators or semiconductors with a direct band gap. Exciton binding energies bear high relevance for industrial applications. It is straightforward to calculate [41] the binding energy of the exciton of the first prominent peak in the optical absorption spectrum and to compare with the experimental one. Table III lists the results from SCAN-mBSE-$f_{xc}^{loc}$ and mTASK-mBSE-$f_{xc}^{loc}$, in comparison with those from TD-DDH, GW-BSE and experiments. In the SCAN-mBSE-$f_{xc}^{loc}$ calculation, a scissor correction to the band gap is used to make the fundamental band gap close to the experimental band gap., For Ar, NaCl and MgO, the experimental binding energies $E_b$ of the first prominent peaks are 2.28, 1.22 and 0.85 eV, respectively. All calculated results listed here underestimate these values, with the exciton binding energies from SCAN-mBSE-$f_{xc}^{loc}$ and mTASK-mBSE-$f_{xc}^{loc}$ close to each other, and close to those from TD-DDH, and approximately close to those from standard GW-BSE.

For monolayers hBN and MoS$_2$, the experimental $E_b$ are 0.70 and 0.58 eV, respectively. In contrast, all the calculated results listed here overestimate these values. Again, the results from GW-BSE, SCAN-mBSE-$f_{xc}^{loc}$ and mTASK-mBSE-$f_{xc}^{loc}$ are approximately or roughly close to each other. The overestimation of the calculated binding energies for monolayer hBN is severer than for monolayer MoS$_2$. These results are not surprising, as hBN is a wide-gap insulator setting a high barrier for theoretical models to capture exciton binding energies. Furthermore, in the monolayer hBN experiment, the monolayer hBN is deposited on a substrate of graphite [39]. The possible substrate screening effect may reduce the measured exciton binding energy more severely, while in monolayer MoS$_2$ experiment [42], the monolayer is free-standing, with less substrate intervened screening effects. Besides, the experimental data listed in Table III for monolayer MoS$_2$ is from a photocurrent measurement [42], which is slightly different from pure optical measurement. Some optical measurements [44] show the position of the first prominent absorption peak of monolayer MoS$_2$ is 1.85 eV. With this correction, the experimental binding energy is 0.65 eV. This brings the results from SCAN-mBSE-$f_{xc}^{loc}$ and mTASK-mBSE-$f_{xc}^{loc}$ even closer to the experimental one, indicating the better accuracy of SCAN-mBSE-$f_{xc}^{loc}$ and mTASK-mBSE-$f_{xc}^{loc}$ for the exciton binding energy calculation of monolayer MoS$_2$ than that of GW-BSE.

## C. Exciton intrinsic lifetimes

Radiative exciton lifetimes impact charge transfer in the photocatalytic process occuring in two-dimensional photocatalysts, making them important characteristics of solar cells [45]. The radiative lifetime of excitons can be derived from Fermi's golden rule with a Hamiltonian describing the interaction between electrons and photons [46]. In a 1D geometrical configuration, where the momentum transfer of photons is along the 1D direction, the intrinsic lifetime of exciton with zero momentum transfer can be calculated as [46]

$$\tau = \frac{1}{\gamma} = \frac{\hbar c^2}{2\pi e^2 \Omega^2} \frac{a}{\mu_a^2}, \tag{4}$$

in the SI unit, where $\gamma$ is the intrinsic radiative decay rate, $\hbar$ Planck constant, $c$ the speed of light, $e$ the electron charge, $\Omega$ the exciton energy, $a$ the length of the unit cell in the 1D tube direction, and $\mu_a^2/a$ the squared exciton transition dipole matrix element per unit tube length. $\mu_a^2 = |\langle 0|z|S\rangle|^2$, where $|0\rangle$ is the ground state wavefunction of the system, $|S\rangle$ the excited state wavefunction with the energy $\Omega$, and $z$ the displacement operator in the 1D direction. The oscillator strength $f_s$ of optical transition is expressed as $f_s = 2|\vec{e} \cdot \langle 0|V|S\rangle|^2/\Omega$ [4], where $\vec{e}$ is the polarization vector of the light and $V$ is the velocity operator. In the 1D case, both $\vec{e}$ and $V$ are along the 1D direction. It can be shown that $\langle 0|V|S\rangle = \langle 0|i[H, \mathbf{r}]|S\rangle = i(E_0 - E_s)\langle 0|\mathbf{r}|S\rangle = -i\Omega\langle 0|\mathbf{r}|S\rangle$. So, $f_s = 2\Omega|\langle 0|\mathbf{r}|S\rangle|^2 = 2\Omega\mu_a^2$. The lifetime of exciton can be written as

$$\tau = \frac{\hbar c^2}{\pi e^2 \Omega} \frac{a}{f_s}. \tag{5}$$

Moreover, Palummo et al. [47] presented a formula of exciton lifetime based on a 2D geometrical configuration, as appears in Eq. (2) in Ref. 47, where the 2D area of the unit cell and the 2D k-point renormalized $\mu_a^2$ are used.

We calculate the exciton intrinsic lifetime with the above Eq. (5) based on 1D, and Eq. (2) of Ref. 47 based on 2D for the six bulk solids (Si, diamond, SiC, Ar, NaCl and MgO) and for the two monolayers hBN and MoS$_2$. Since a 3D material can be treated as a very thick 2D slab, we believe that the 2D based formula can be more suitable and transferable to 3D (three-dimensional) materials. However, the 1D based formula may be also applicable to highly anisotropic 2D or 3D materials.

Figure 3 shows the results from SCAN-mBSE-$f_{xc}^{loc}$. The plotted panels on the left column in Figure 3 are calculated with the above Eq. (5), while the right column is with the Eq. (2) of Ref. 47. For Si, SiC, and MoS$_2$, the lifetime calculated from the 1D formula is about one order of magnitude higher than that from the 2D formula. For diamond, Ar, NaCl, MgO and hBN, the results from the two formulas are about the same order of magnitude. Generally speaking, the exciton lifetimes from the 2D formula for most of the bright excitons of the eight materials considered here are about several nanoseconds.

Figure 4 shows the results calculated with mTASK-mBSE-$f_{xc}^{loc}$ with both the 1D and 2D formulas. The overall trend in the results from mTASK-mBSE-$f_{xc}^{loc}$ resembles that from SCAN-mBSE-$f_{xc}^{loc}$. For Si, SiC, and MoS2, the lifetime calculated from the 1D formula is about one order of magnitude higher than that from the 2D formula. For diamond, Ar, NaCl, MgO and hBN, the results from the two formulas are about the same order of magnitude. Also, the exciton lifetimes from the 2D formula for most of the bright excitons of the eight materials considered here are about several nanoseconds, agreeing with those from SCAN-mBSE-$f_{xc}^{loc}$, indicating the approximate consistency of the lifetime calculations from SCAN-mBSE-$f_{xc}^{loc}$ and mTASK-mBSE-$f_{xc}^{loc}$. Note that that the intrinsic exciton lifetime cannot be directly measured in experiments, where phonon coupling and temperature effects are involved [47].

## IV. CONCLUSIONS

In conclusion, we combine advanced metaGGA functionals, SCAN and mTASK, with a dielectric function model based BSE (mBSE) method to calculate the optical response properties of bulk and monolayer solid materials. The model dielectric function employed in this method can capture the screening effect in materials reasonably well with computation affordability. The calculated optical absorption spectra (the imaginary part of the macroscopic dielectric function) of bulk Si, diamond, SiC, MgO, and monolayer MoS$_2$ are qualitatively agree with the experimental spectra. For the large-gap insulating solids Ar and NaCl, the sharp absorption double peak in the optical absorption curves is found to be challenging to reproduce. Although with difference in details, the optical absorption curves calculated with SCAN-mBSE-$f_{xc}^{loc}$ are qualitatively similar or close to those from the more computationally demanding TD-DDH and standard GW-BSE. The exciton binding energies of the first prominent peak in the optical absorption spectra of the direct band gap solids Ar, NaCl and MgO calculated with SCAN-mBSE-$f_{xc}^{loc}$ and mTASK-mBSE-$f_{xc}^{loc}$ qualitatively agree with those from TD-DDH and GW-BSE, and moderately underestimate the experimental values. For the large-gap insulating monolayer hBN, GW-BSE, SCAN-mBSE-$f_{xc}^{loc}$ and mTASK-mBSE-$f_{xc}^{loc}$ all overestimate the binding energy of the first prominent peak. For monolayer $MoS_2$, SCAN-mBSE-$f_{xc}^{loc}$ and mTASK-mBSE-$f_{xc}^{loc}$ give the binding energy of the first prominent peak close to the experimental value, better than GW-BSE does. The calculated intrinsic exciton lifetimes with SCAN-mBSE-$f_{xc}^{loc}$ and mTASK-mBSE-$f_{xc}^{loc}$ for the eight materials show a consistent agreement, with magnitudes of several nanoseconds for most bright excitons. Overall, the SCAN-mBSE-$f_{xc}^{loc}$ and mTASK-mBSE-$f_{xc}^{loc}$ methods provide a computationally efficient alternative to calculate the optical properties of materials, with an approximately qualitative accuracy. Further improvements can be expected by including more spatial nonlocality in the exchange-correlation kernel part $f_{xc}^{loc}$.

## ACKNOWLEDGEMENTS

This work was supported by the donors of ACS Petroleum Research Fund under New Directions Grant 65973-ND10. A.R. served as Principal Investigator on ACS PRF 65973-ND10 that provided support for H.T. A.R. acknowledges support from Tulane University's startup fund. This research includes calculations carried out on the high-performance computing (HPC) resources at NERSC supported in part by the National Science Foundation through major research instrumentation grant number 1625061 and by the US Army Research Laboratory under contract number W911NF-16-2-0189. This research was supported in part by HPC resources and services provided by Information Technology at Tulane University, New Orleans, LA.

# References

1. P. Ondračka, D. Holec, L. Zajíčková, Optical Properties from Ab Initio Calculations, Optical Characterization of Thin Solid Films, edited by O. Stenzel, M. Ohlídal, (Springer, Cham, 2018).
2. G. Liu, Y. Shang, B. Jia, X. Guan, L. Han, X. Zhang, H. Song, and P. Lu, First-principles calculation of the optical properties of the $YBa_2Cu_3O_{7-\delta}$ oxygen vacancies model, RSC Adv. **13,** 18927 (2023).
3. Y. Rouzhahong, M. Wushuer, M. Mamat, Q. Wang, and Q. Wang, First Principles Calculation for Photocatalytic Activity of GaAs Monolayer, Scientific Reports **10,** 9597 (2020).
4. J. Deslippe, G. Samsonidze, D. A. Strubbe, M. Jain, M. L. Cohen, and S. G. Louie, BerkeleyGW: A Massively Parallel Computer Package for the Calculation of the Quasiparticle and Optical Properties of Materials and Nanostructures, Comput. Phys. Commun. **183,** 1269 (2012).
5. M. Rohlfing and S. G. Louie, Electron-Hole Excitations and Optical Spectra from First Principles, Phys. Rev. B **62,** 4927 (2000).
6. L. Reining, The GW approximation: content, successes and limitations, Advanced Review **8,** 1344 (2018).
7. G. Onida, L. Reining, and A. Rubio, Electronic excitations: Density-functional versus many-body Green's-function approaches, Rev. Mod. Phys. **74,** 601 (2002).
8. E. K. U. Gross and W. Kohn, Local Density-Functional Theory of Frequency-Dependent Linear Response, Phys. Rev. Lett. **55,** 2850 (1985).
9. E. Runge and E. K. U. Gross, Density-Functional Theory for Time-Dependent Systems, Phys. Rev. Lett. **52,** 997 (1984).
10. S. Sharma, J.K. Dewhurst, A. Sanna, and E.K.U. Gross, Phys. Rev. Lett. **107**, 186401 (2011).
11. S. Botti, F. Sottile, N. Vast, V. Olevano, L. Reining, H.-C. Weissker, A. Rubio, G. Onida, R. Del Sole, and R. W. Godby, Long-range contribution to the exchange-correlation kernel of time-dependent density functional theory, Phys. Rev. B **69**, 155112 (2004).
12. F. Sottile, V. Olevano, and L. Reining, Parameter-free Calculation of Response Functions in Time-Dependent Density Functional Theory, Phys. Rev. Lett. **91**, 056402 (2003).
13. A. Marini, R. Del Sole, and A. Rubio, Bound Excitons in Time Dependent Density-Functional Theory: Optical and Energy Loss Spectra, Phys. Rev. Lett. **91**, 256402 (2003).
14. W. Chen, G. Miceli, G.M. Rignanese, and A. Pasquarello, Nonempirical dielectric-dependent hybrid functional with range separation for semiconductors and insulators, Phys. Rev. Mater. **2**, 073803 (2018).
15. A. Tal, P. Liu, G. Kresse, and A. Pasquarello, Accurate optical spectra through time-dependent density functional theory based on screening-dependent hybrid functionals, Phys. Rev. Research **2**, 032019(R) (2020).

16. J. Sun, A. Ruzsinszky, and J. P. Perdew, Strongly Constrained and Appropriately Normed Semilocal Density Functional, Phys. Rev. Lett. **115**, 036402 (2015).
17. B. Neupane, H. Tang, N. K. Nepal, S. Adhikari, and A. Ruzsinszky, Opening band gaps of low-dimensional materials at the meta-GGA level of density functional approximations, Phys. Rev. Materials **5**, 063803 (2021).
18. W. Kohn and L. J. Sham, Self-Consistent Equations Including Exchange and Correlation Effects, Phys. Rev. **140**, A1133 (1965).
19. J. P. Perdew, K. Burke, and M. Ernzerhof, Generalized Gradient Approximation Made Simple, Phys. Rev. Lett. **77**, 3865 (1996).
20. V. U. Nazarov and G. Vignale, Optics of Semiconductors from Meta-Generalized-Gradient-Approximation-Based Time-Dependent Density-Functional Theory, Phys. Rev. Lett. **107**, 216402 (2011).
21. G. Kresse and J. Furthmüller, Efficient Iterative Schemes for Ab Initio Total-Energy Calculations Using a Plane-Wave Basis Set, Phys. Rev. B **54**, 11169 (1996).
22. G. Kresse and D. Joubert, From Ultrasoft Pseudopotentials to the Projector Augmented-Wave Method, Phys. Rev. B **59**, 1758 (1999).
23. R. Peverati and D. G. Truhlar, Performance of the M11-L density functional for bandgaps and lattice constants of unary and binary semiconductors, J. Chem. Phys. **136**, 134704 (2012).
24. C. S. Barrett and L. Meyer, X-Ray Diffraction Study of Solid Argon, J. Chem. Phys. 41, 1078–1081 (1964)
25. D. B. Sirdeshmukh, L. Sirdeshmukh, and K. G. Subhadra, Alkali Halides: A Handbook of Physical Properties (Springer Berlin, 2001).
26. I. Fongkaew, B. Yotburut, W. Sailuam, W. Jindata, T. Thiwatwaranikul, A. Khamkongkaeo, N. Chuewangkam, N. Tanapongpisit, W. Saenrang, R. Utke, P. Thongbai, S. Pinitsoontorn, S. Limpijumnong, and W. Meevasana, Effect of hydrogen on magnetic properties in MgO studied by first-principles calculations and experiments, Scientific Reports **12**, 10063 (2022).
27. W. Auwarter, T. J. Kreutz, T. Greber, J. Osterwalder, XPD and STM investigation of hexagonal boron nitride on Ni(111), Surface Science **429** 229-236 (1999).
28. Y. L. Huang, Y. Chen, W. Zhang, S. Y. Quek, C.-H. Chen, L.-J. Li, W.-T. Hsu, W.-H. Chang, Y. J. Zheng, W. Chen, and A. T. S. Wee, Bandgap tunability at single-layer molybdenum disulphide grain boundaries, Nature Communications **6**, 6298 (2015).
29. M. E. Casida, Time-dependent density-functional response theory for molecules, Recent Advances in Density Functional Methods, Part I, edited by D. P. Chong, (World Scientific, Singapore, 1995).
30. A. M. Ferrari, R. Orlando, and M. Rerat, Ab Initio Calculation of the Ultraviolet–Visible (UV-vis) Absorption Spectrum, Electron-Loss Function, and Reflectivity of Solids, J. Chem. Theory Comput. **11**, 3245 (2015).

31. P. Lautenschlager, M. Garriga, L. Vina, and M. Cardona, Temperature dependence of the dielectric function and interband critical points in silicon, Phys. Rev. B **36**, 4821 (1987).
32. E. D. Palik, Handbook of Optical Constants of Solids, Vol.1 (Academic, New York, 2012).
33. S. Logothetidis and J. Petalas, Dielectric function and reflectivity of 3C-silicon carbide and the component perpendicular to the c axis of 6H-silicon carbide in the energy region 1.5–9.5 eV, J. Appl. Phys. **80**, 1768 (1996).
34. V. Saile, M. Skibowski, W. Steinmann, P. Gürtler, E. E. Koch, and A. Kozevnikov, Observation of Surface Excitons in Rare Gas Solids, Phys. Rev. Lett. **37**, 305 (1976).
35. D. M. Roessler and W. C. Walker, Electronic spectra of crystalline NaCl and KCl, Phys. Rev. **166**, 599 (1968).
36. M. L. Bortz, R. H. French, D. J. Jones, R. V. Kasowski, and F. S. Ohuchi, Temperature dependence of the electronic structure of oxides: MgO, $MgAl_2O_4$ and $Al_2O_3$, Phys. Scr. **41**, 537 (1990).
37. K. F. Mak, C. Lee, J. Hone, J. Shan, and T.F. Heinz, Atomically Thin $MoS_2$: A New Direct-Gap Semiconductor, Phys. Rev. Lett. **105**, 136805 (2010).
38. R. J. Hunt, B. Monserrat, V. Zólyomi, and N. D. Drummond, Diffusion quantum Monte Carlo and GW study of the electronic properties of monolayer and bulk hexagonal boron nitride, Phys. Rev. B **101**, 205115 (2020).
39. R. J. P. Roman, F. J. R. Costa Costa, A. Zobelli, C. Elias, P. Valvin, G. Cassabois, B. Gil, A. Summerfield, T. S. Cheng, C. J. Mellor, P. H. Beton, S. V. Novikov, and L. F. Zagonel, Band gap measurements of monolayer h-BN and insights into carbon-related point defects, 2D Mater. **8**, 044001 (2021).
40. D. Y. Qiu, F. H. da Jornada, and S. G. Louie, Optical Spectrum of $MoS_2$: Many-Body Effects and Diversity of Exciton States, Phys. Rev. Lett. **111**, 216805 (2013).
41. M. Combescot and S.-Y. Shiau, Excitons and Cooper Pairs: Two Composite Bosons in Many-Body Physics (Oxford University Press, 2016).
42. A. R. Klots, A. K. M. Newaz, B. Wang, D. Prasai, H. Krzyzanowska, J. Lin, D. Caudel, N. J. Ghimire, J. Yan, B. L. Ivanov, K. A. Velizhanin, A. Burger, D. G. Mandrus, N. H. Tolk, S. T. Pantelides, and K. I. Bolotin, Probing excitonic states in suspended two-dimensional semiconductors by photocurrent spectroscopy, Scientific Reports **4**, 6608 (2014).
43. L. Wirtz, A. Marini, and A. Rubio, A Excitons in boron nitride nanotubes: dimensionality effects, Phys. Rev. Lett. **96** 126104 (2006).
44. A. Splendiani, L. Sun, Y. Zhang, T. Li, J. Kim, C.-Y. Chim, G. Galli, and F. Wang, Emerging Photoluminescence in Monolayer $MoS_2$, Nano Lett. **10**, 1271 (2010).
45. W. Xie, L. Tian, K. Wu, B. Guo, and J. R. Gong, Understanding and modulating exciton dynamics of organic and low-dimensional inorganic materials in photo(electro)catalysis, J. Catalysis **395**, 91 (2021).

46. C. D. Spataru, S. Ismail-Beigi, R. B. Capaz, and S. G. Louie, Theory and Ab Initio Calculation of Radiative Lifetime of Excitons in Semiconducting Carbon Nanotubes, Phys. Rev. Lett. **95**, 247402 (2005).
47. M. Palummo, M. Bernardi, and J. C. Grossman, Exciton Radiative Lifetimes in Two-Dimensional Transition Metal Dichalcogenides, Nano Lett. **15**, 2794 (2015).

Tables

Table I. The screening parameters $\alpha$ and $\mu$ obtained by fitting the inverse dielectric function calculated with $G_0W_0$@SCAN to Eq. (1) for six bulk solids Si, diamond, SiC, Ar, NaCl and MgO, and two 2D monolayers hBN and $MoS_2$. $\alpha$ is dimensionless and the unit of $\mu$ is 1/Å.

|   | Si | Diamond | SiC | Ar | NaCl | MgO | hBN | $MoS_2$ |
|---|---|---|---|---|---|---|---|---|
| $\alpha$ | 0.09 | 0.19 | 0.17 | 0.60 | 0.46 | 0.34 | 0.60 | 0.28 |
| $\mu$ | 0.65 | 1.50 | 1.25 | 1.08 | 1.05 | 1.18 | 0.80 | 0.80 |

Table II. The calculated band gaps of six bulk solids and two 2D monolayer materials with different methods, in comparison with experimental values. Values are in eV. Entries marked with "-" mean no data available.

|  | Exp. | TD-DDH | GW | SCAN | mTASK |
|---|---|---|---|---|---|
| Si | 1.23[a] | 1.31[a] | 1.41[a] | 0.82 | 1.33 |
| Diamond | 5.85[a] | 5.69[a] | 5.85[a] | 4.55 | 4.53 |
| SiC | 2.53[a] | 2.50[a] | 2.55[a] | 1.43 | 2.03 |
| Ar | 14.33[a] | 14.60[a] | 13.75[a] | 9.52 | 14.87 |
| NaCl | 9.14[a] | 9.13[a] | 8.86[a] | 6.05 | 9.96 |
| MgO | 8.36[a] | 8.41[a] | 8.12[a] | 5.56 | 7.87 |
| hBN | 6.80[b] | - | 8.20[c] | 4.82 | 5.93 |
| $MoS_2$ | 2.50[d] | - | 2.84[e] | 1.60 | 1.89 |

[a] Reference 15.
[b] Reference 39.
[c] Reference 43.
[d] Reference 42.
[e] Reference 40.

Table III. The calculated direct band gaps $E_g$, exciton energies of the first prominent peak in the optical absorption $E_{ext}$ and exciton binding energies $E_b$ of three bulk solids and two 2D monolayer materials in comparison with experimental values. $E_b = E_g - E_{ext}$ with all units in eV. The $E_g$ values from SCAN-mBSE-$f_{xc}^{loc}$ are scissor-corrected and close to the experimental values.

|  |  | Ar | NaCl | MgO | hBN | MoS$_2$ |
|---|---|---|---|---|---|---|
| Exp. | $E_g$ | 14.33[a] | 9.14[a] | 8.36[a] | 6.80[b] | 2.50[c] |
|  | $E_{ext}$ | 12.05[d] | 7.92[d] | 7.51[d] | 6.10[b] | 1.92[c] |
|  | $E_b$ | 2.28 | 1.22 | 0.85 | 0.70 | 0.58 |
|  |  |  |  |  |  |  |
| TD-DDH | $E_g$ | 14.60[a] | 9.13[a] | 8.41[a] | - | - |
|  | $E_{ext}$ | 12.79[d] | 8.30[d] | 8.19[d] | - | - |
|  | $E_b$ | 1.81 | 0.83 | 0.22 | - | - |
|  |  |  |  |  |  |  |
| GW-BSE | $E_g$ | 13.75[a] | 8.86[a] | 8.12[a] | 8.20[e] | 2.84[f] |
|  | $E_{ext}$ | 11.73[d] | 7.89[d] | 7.79[d] | 6.10[g] | 1.88[f] |
|  | $E_b$ | 2.02 | 0.97 | 0.33 | 2.10 | 0.96 |
|  |  |  |  |  |  |  |
| SCAN-mBSE-$f_{xc}^{loc}$ | $E_g$ | 14.33 | 9.14 | 8.36 | 6.10 | 2.60 |
|  | $E_{ext}$ | 12.84 | 8.39 | 7.91 | 3.41 | 1.87 |
|  | $E_b$ | 1.49 | 0.75 | 0.45 | 2.69 | 0.73 |
|  |  |  |  |  |  |  |
| mTASK-mBSE-$f_{xc}^{loc}$ | $E_g$ | 14.87 | 9.96 | 7.87 | 5.93 | 1.89 |
|  | $E_{ext}$ | 13.30 | 9.13 | 7.60 | 4.05 | 1.20 |
|  | $E_b$ | 1.57 | 0.83 | 0.27 | 1.88 | 0.69 |

[a] Reference 15.
[b] Reference 39.
[c] Reference 42.
[d] determined from the curves in Reference 15.
[e] Reference 43.
[f] Reference 40.
[g] Reference 38.

Figures

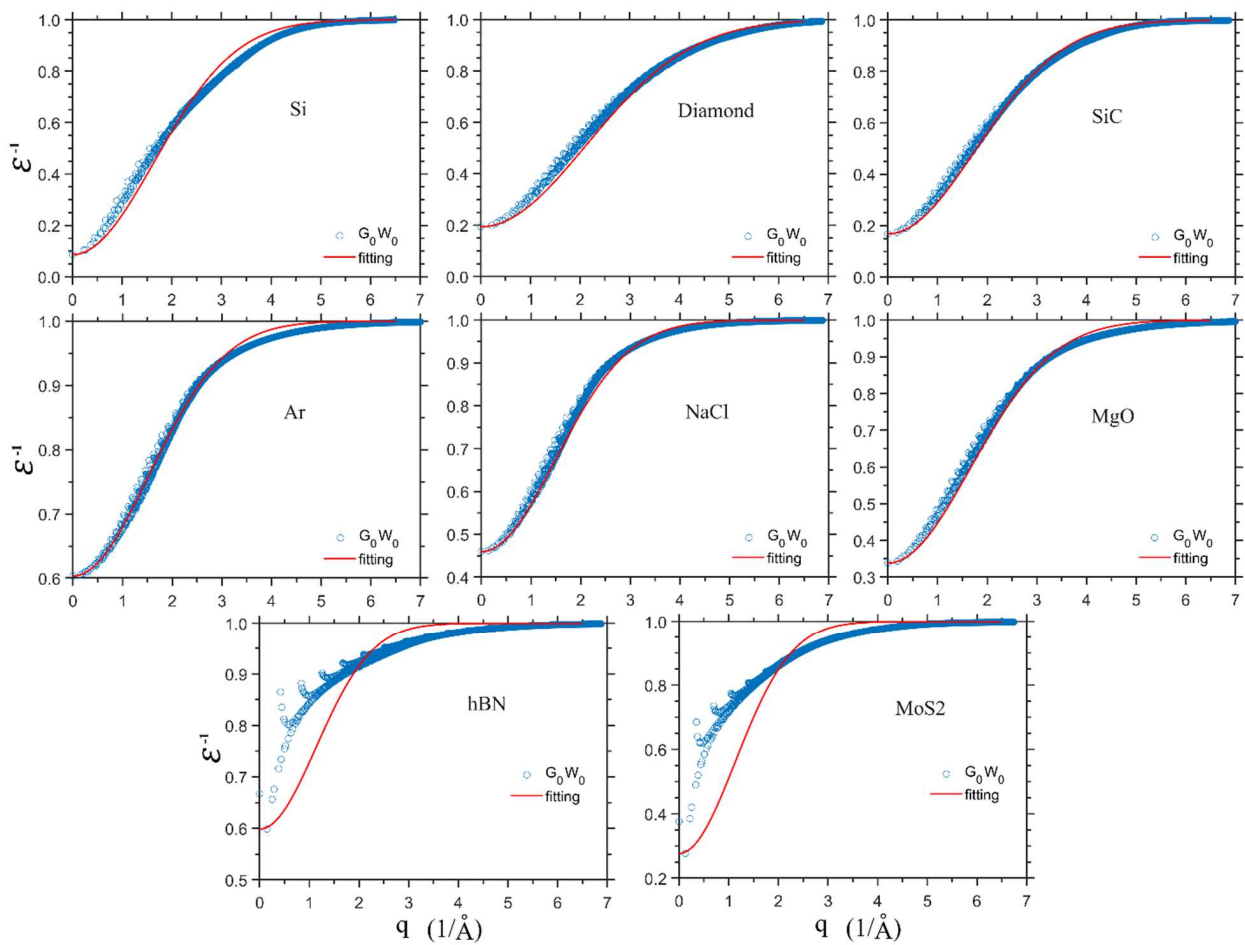

FIG. 1. The inverse of the static dielectric function as a function of wave vector calculated with $G_0W_0$@SCAN and the fitting to the dielectric model of Eq. (1) for six bulk solids Si, diamond, SiC, Ar, NaCl and MgO, and two 2D monolayers hBN and $MoS_2$.

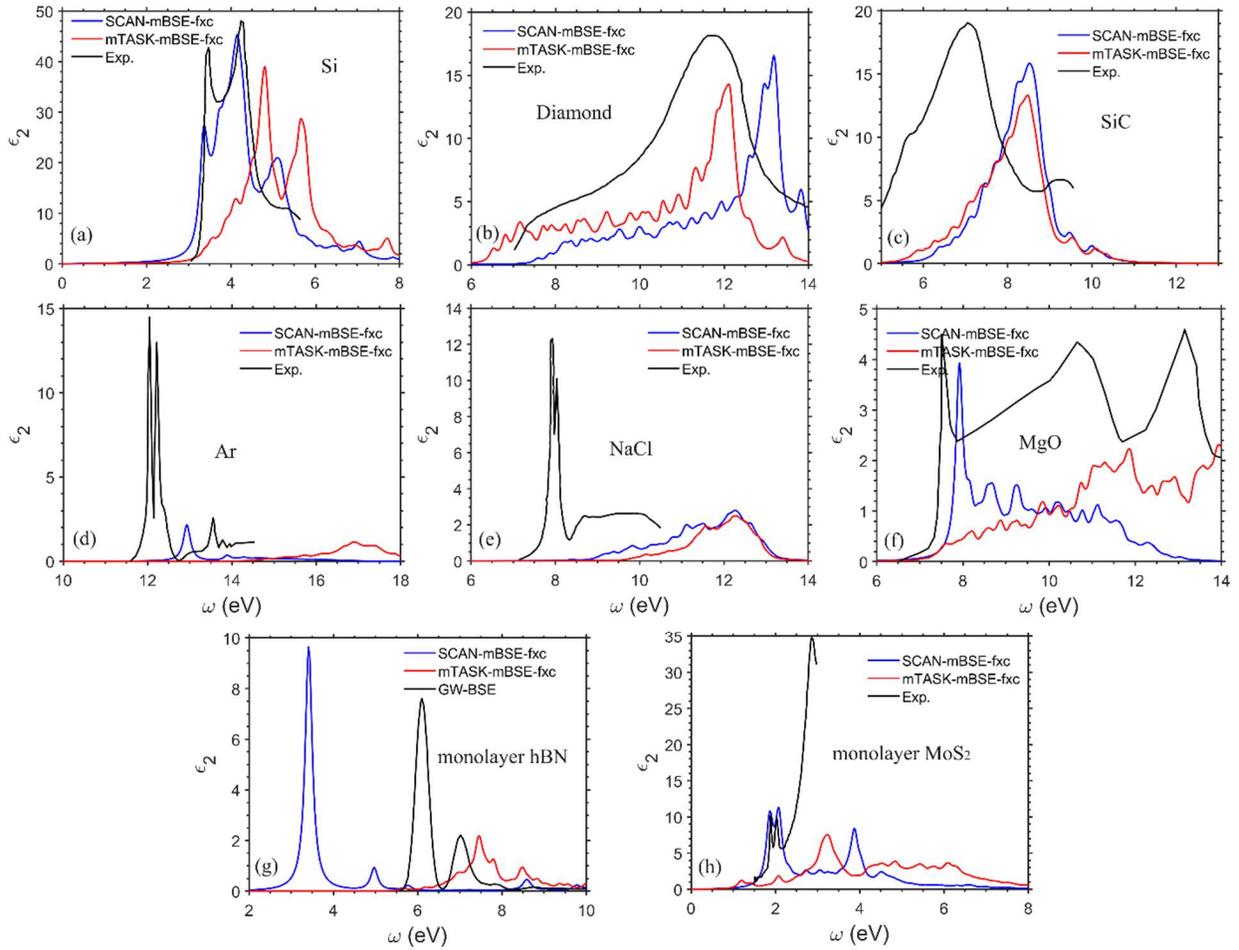

FIG. 2. The optical absorption spectra calculated with SCAN-mBSE-$f_{xc}^{loc}$ and mTASK-mBSE-$f_{xc}^{loc}$ in comparison with the experimental or GW-BSE for bulk Si, diamond, SiC, Ar, NaCl and MgO, and 2D monolayers hBN and MoS$_2$. The optical absorption is represented by the imaginary part of the calculated macroscopic dielectric function. The experimental data are taken from Ref. 31 for Si, Ref. 32 for diamond, Ref. 33 for SiC, Ref. 34 for Ar, Ref. 35 for NaCl, Ref. 36 for MgO, and Ref. 37 for monolayer MoS$_2$. The GW-BSE data for monolayer hBN is from Ref. 38.

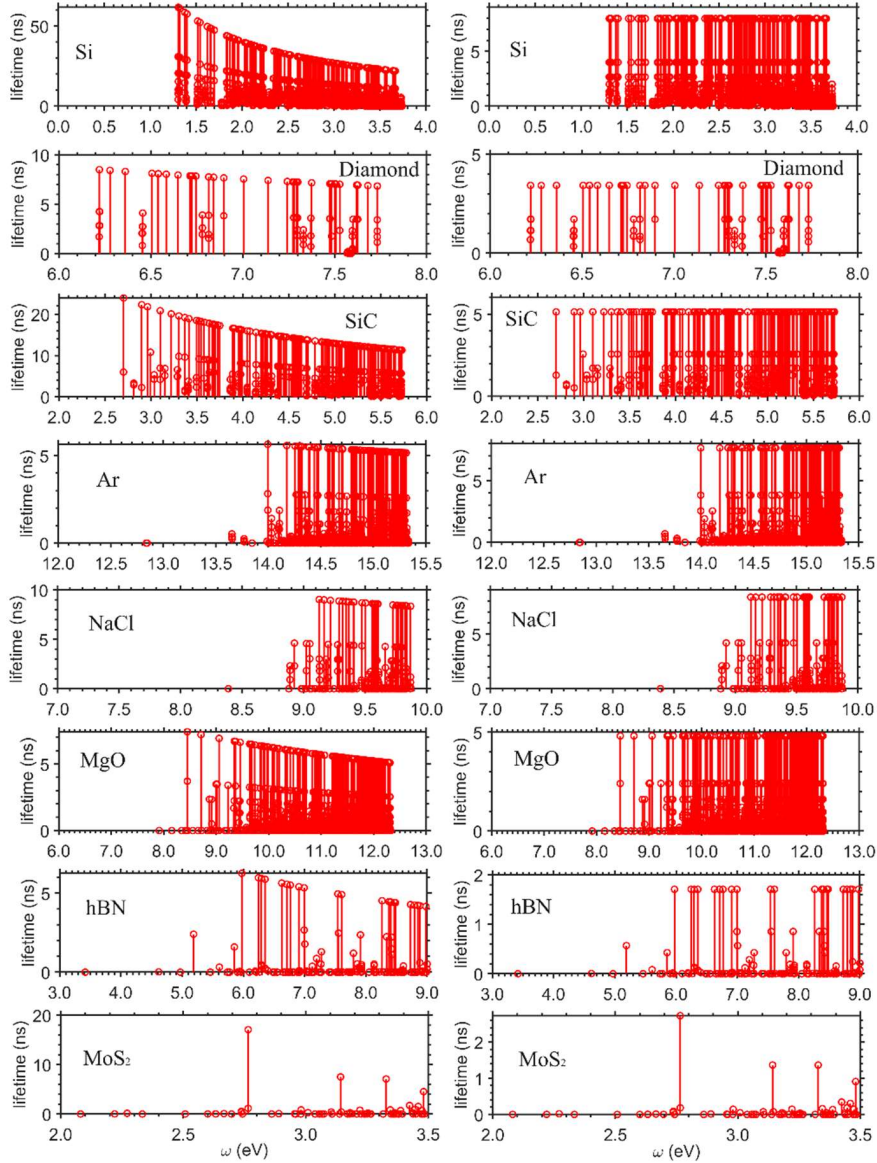

FIG. 3. The calculated exciton lifetimes of six bulk solids (Si, diamond, SiC, Ar, NaCl and MgO) and two monolayers hBN and $MoS_2$ from the SCAN-mBSE-$f_{xc}^{loc}$ method. For each material, the left panel plot is calculated from the formula based on a one-dimensional geometry treatment and derivation, while the right one is from the formula based on two-dimensional geometry treatment and derivation.

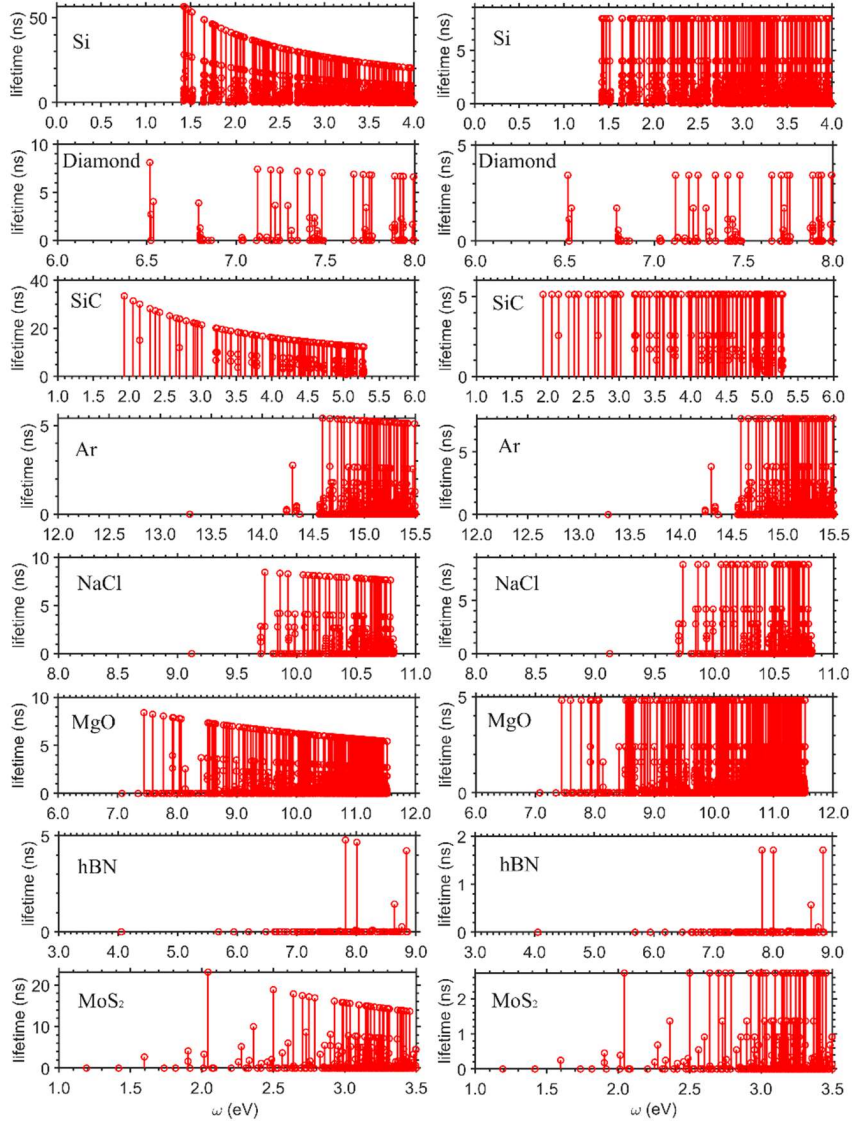

FIG. 4. The calculated exciton lifetimes of the eight materials from the mTASK-mBSE-$f_{xc}^{loc}$ method. The panels are plotted and arranged in the same way as in Figure 3.